\documentclass[aps,prb,superscriptaddress,twocolumn,floatfix,showpacs,a4paper]{revtex4-1}

\usepackage{amsfonts,amsmath,amssymb}
\usepackage{textcomp}
\usepackage{bm}
\usepackage{upgreek}
\usepackage{graphicx}
\usepackage[latin1]{}
\usepackage{color}

\begin{document}
\title{Spin Manipulation in Graphene \\
by Chemically-Induced Sublattice Pseudospin Polarization}

\author{Dinh Van Tuan} 
\affiliation{Catalan Institute of Nanoscience and Nanotechnology (ICN2), CSIC and The Barcelona Institute of Science and Technology, Campus UAB, Bellaterra, 08193 Barcelona, Spain }
\author{Stephan Roche}
\affiliation{Catalan Institute of Nanoscience and Nanotechnology (ICN2), CSIC and The Barcelona Institute of Science and Technology, Campus UAB, Bellaterra, 08193 Barcelona, Spain }
\affiliation{ICREA, Instituci\'{o} Catalana de Recerca i Estudis Avan\c{c}ats, 08070 Barcelona, Spain}
\email{stephan.roche@icn.cat}
\date{\today}

\begin{abstract}
Spin manipulation is one of the most critical challenges to realize spin-based logic devices and spintronic circuits. Graphene has been heralded as an ideal material to achieve spin manipulation but so far new paradigms and demonstrators are limited. Here we show that certain impurities such as fluorine ad-atoms, which locally break sublattice symmetry without the formation of strong magnetic moment, could result in a remarkable variability of spin transport characteristics. The impurity resonance level is found to be associated with a long range sublattice pseudospin polarization, which by locally decoupling spin and pseudospin dynamics, provokes a huge spin lifetime electron-hole asymmetry. In the dilute impurity limit, spin lifetimes could be tuned electrostatically from hundred picoseconds to several nanoseconds, providing a protocol to chemically engineer an unprecedented spin device functionality.
\end{abstract}


\maketitle

{\it Introduction.} The possibility to fine tune the electronic, charge and spin transport properties of graphene using chemical functionalization \cite{Neto2009,Krasheninnikov2007,KPL2010,Craciun2013}, irradiation (defect formation) \cite{Nair2012}, electric fields \cite{Son2006} or antidot fabrication \cite{Pedersen2008} has become an exciting field of research with almost endless possibilities. In particular, chemical treatments such as ozonisation, hydrogenation, or fluorination, introducing a variable density of surface ad-atoms from typically 0.001\% to few percent, have demonstrated a large spectrum of accessible physical states from anomalous transport to highly insulating behavior of chemically reactive graphene derivatives \cite{Elias2009,Withers2010,Moser2010,Cheng2010,Robinson2010}. On the other hand, graphene  exhibits long room temperature spin lifetime and rich surface chemistry opportunities which could be harnessed for the development of all-spin logic technologies \cite{Tombros2007,Dlubak2012,Seneor2012,Dery2012,Folk2013,Kamalakar2015,Kawakami2015,Roche2015}. As a matter of illustration, the use of chemical fluorination of graphene bilayer has been shown to yield very high spin injection efficiency (above $60\%$), owing to improved interface spin filtering \cite{Friedman2014}.

Spin lifetime is an essential quantity that fixes the upper time and length scales on which spin devices can operate, so that knowing its value and variability are prerequisite to realizing graphene spintronic technologies. The sources of spin relaxation turn out to be diversified in graphene, and extrinsic disorder driven by ad-atom impurities can significantly enhanced spin-orbit interaction around defects \cite{Neto2009}, or create local magnetism \cite{Yazyev2008}; both effects usually reducing spin lifetimes, even in the dilute limit \cite{Huertas-Hernando2009, Ochoa2012, Kochan2014, Soriano2015, Kochan2015,Omar2015, Thomsen2015,Bundesmann2015,Folk2015}.

The nature of spin relaxation in graphene has been initially discussed either in terms of Elliot-Yafet \cite{Ochoa2012} or Dyakonov-Perel \cite{Ertler2009} mechanisms, depending on the scaling of spin lifetime with defect density. However recently, a novel spin relaxation mechanism in non-magnetic graphene samples has been connected to the unique spin-pseudospin entanglement occurring near the Dirac point, pointing towards revisiting the role of sublattice pseudospin \cite{Dinh2014}. 

Sublattice pseudospin is an additional quantum degree of freedom, mathematically very similar to spin and unique to graphene sublattice degeneracy \cite{Neto2009}. In absence of spin-orbit coupling, the low-energy electronic states write $\Psi_{\vec{k}}(\vec{r}) \sim (\psi_{A} (1,0)^{\rm T} +\psi_{B} (0,1)^{\rm T})\times {\rm e}^{i\vec{k}\vec{r}}$, where $(1,0)^{\rm T}$ and $(0,1)^{\rm T}$ define up and down-pseudospin states, while $\psi_{A}$ (resp. $\psi_{B}$) give the wavefunction weight restricted to A (resp. B) sublattice sites \cite{Neto2009,Liu2011}. In addition to sublattice pseudospin, valley isospin (for the two K-points in the reciprocal space) also shows up  in  the electronic wavefunctions, and harnessing these degrees of freedom is the target of {\it "Valleytronics and Pseudospintronics"} \cite{Pesin2012,Zuelicke}.  The complex interplay between sublattice pseudospin and valley isospin is currently the source of innovative device proposals such as valley or pseudospin filtering and switches \cite{Son2006,Rycerz2007,SanJose2009, Tkachov2009,Park2011,Gunlycke2011,Lundeberg2014}.

In presence of a Rashba spin-orbit coupling (SOC) field either generated by a substrate-induced electric field or a weak density of metal ad-atoms (gold, nickel), spin and pseudospin become strongly coupled at the Dirac point where the eigenstates take the form $\Psi_{\vec{k}=\vec{K}} \sim (1,0)^{\rm T}\times|\downarrow\rangle \pm i (0,1)^{\rm T} \times |\uparrow\rangle$, where $|\downarrow\rangle$ and $|\uparrow\rangle$ denote the spin state \cite{Rashba2009,Dinh2014}.  Such spin-pseudospin locking drives to an entangled dynamics of spin and pseudospin  resulting in fast spin dephasing, even when approaching the ballistic limit \cite{Dinh2014}, with increasing spin lifetimes away from the Dirac point, as observed experimentally \cite{Guimaraes2014}. This phenomenon suggests ways to engineer spin manipulation based on controlling the pseudospin degree of freedom (or vice versa), which would help in the development of spin logics \cite{Dery2012,Kawakami2015,Roche2015}.

In this Letter, we reveal that chemical functionalization of graphene with certain types of ad-atoms such as fluorine, by breaking the sublattice symmetry and by inducing a SOC without the formation of strong magnetic moment, provide an enabling technique to monitor spin transport properties in a remarkable way for spintronic applications. The fluorine ad-atoms indeed produce hole impurity levels which exhibit a long range spatial sublattice pseudospin polarization (SPP), which counteracts the homogeneous Rashba SOC field at the origin of the intrinsic spin precession and relaxation in the otherwise fluorine free samples \cite{DinhScRep}. As a result, spin and pseudospin dynamics are not anymore coupled at the impurity resonances, which lead to the possibility to electrostatically tune spin lifetime by up to one order of magnitude (for instance under electrostatic gating). This is a theoretical opportunity for designing a new kind of spin transistor effect based on a gate-controlled spin transport length. Calculations are performed using a realistic tight-binding model elaborated from ab-initio calculations, whereas the spin dynamics is computed through the time-evolution of the expectation value of the spin operator projected on a real space basis set. 

{\it Tight-binding description of fluorinated graphene.} The description of fluorine ad-atom on graphene is achieved using a tight-binding model elaborated from ab-initio simulations \cite{Irmer2015}. The Hamiltonian for the system involves two parts:
\begin{equation}
{\mathcal{H}}={\mathcal{H}}_G+{\mathcal{H}}_{FG}
\label{Hamil}
\end{equation}
The first part describes the graphene in a homogeneous SOC field induced by the substrate or gate voltage
\begin{eqnarray}
{\mathcal{H}}_G&=&-\gamma_0\sum_{\langle ij\rangle }c_i^+c_j+\frac{2i}{9}\lambda_{I}\sum_{\langle\langle ij\rangle\rangle}c_i^+\vec{s}\cdot(\vec{d}_{kj}\times\vec{d}_{ik})c_j
\nonumber\\
&+&\frac{2i}{3}\lambda_R\sum_{\langle ij\rangle }c_i^+   \vec{z}\cdot(\vec{s}\times\vec{d}_{ij})c_j   
\label{HamilG}
\end{eqnarray}
where $\gamma_0$ is the usual $\pi$-orbital hopping term between nearest-neighbors,  $\lambda_I=12\ \mu$eV is commonly value used for the intrinsic SOC of graphene \cite{Gmitra2009} while the Rashba SOC $\lambda_R$ is an electric field-dependent quantity. In this study we take $\lambda_R=37.4\ \mu$eV taken from an extended $sp$-band tight-binding model \cite{Ast2012} for graphene under the influence of an electric field of $0.1\ V/$\AA~, induced by the substrate or the gate voltage. 

The second part $\hat{\mathcal{H}}_{FG}$ describes the influences of fluorine on graphene
\begin{eqnarray}
 \hat{\mathcal{H}}_{FG} &=&\epsilon_F\sum_{m}F_m^+F_m +T\sum_{\langle mi\rangle }\left[F_m^+A_i+h.c.\right]  \nonumber\\
&+& \frac{2i}{9}\Lambda_{I}^B\sum_{\langle\langle ij\rangle\rangle }B_i^+\vec{s}.(\vec{d}_{kj}\times\vec{d}_{ik})B_j     \nonumber\\              
 &+& \frac{2i}{3}\Lambda_{R}\sum_{\langle ij\rangle}\left[A_i^+\vec{\hat{z}}.(\vec{s}\times\vec{d}_{ij})B_j+h.c.\right]   \nonumber\\
&+& \frac{2i}{3}\Lambda_{PIA}\sum_{\langle\langle ij\rangle\rangle}B_i^+\vec{\hat{z}}.(\vec{s}\times\vec{d}_{ij})B_j  
 \label{HamilFG}
\end{eqnarray}
with all the parameters $\epsilon_F=-2.2$ eV, $T=5.5$ eV, $\Lambda_I^B=3.3$ meV, $\Lambda_R=11.2$ meV, and $\Lambda_{PIA}^B=7.3$ meV are derived from ab-initio simulations  \cite{Irmer2015}. The operator $F$ ($F^+$) annihilates (creates) an electron in the atomic $p_z$ orbital on fluorine F. $A$ and $B$ ($A^+$ and $B^+$ ) denote the annihilation (creation)
operators for $p_z$ orbital on fluorinated carbons and their nearest neighbors, respectively.  The first term in above Hamiltonian is the on-site energy term on the fluorine ad-atoms and the second term is the hopping term between fluorine ad-atoms $F$ and fluorinated carbon $A\equiv C_F$. The third and the fourth terms  which are similar to the SOC terms in Eq.(\ref{HamilG}) simulate the local intrinsic and Rashba SOCs induced by the absorption of fluorine on graphene. Finally the last term, the new SOC term, coming from the pseudospin inversion asymmetry (PIA) mediates the  spin-flip hopping between two second nearest neighbors $B_i$. It is worth mentioning that we are using the $\pi$-orbital tight binding model which is different from a recent paper on  the electronic structures and optical properties of fluorinated graphene in which the multi-orbital tight-binding was employed \cite{Shengjun2015}.

 {\it Spin dynamics methodology.} The spin dynamics of electron in fluorinated graphene is investigated using the time-dependent evolution of the spin polarization of propagating wavepackets \cite{Dinh2014}. Simulations of samples of $\mu{\rm m}^{2}$ size are performed, containing hundred millions of carbon atoms ($N\sim 10^{8}$). The time-evolution of the spin polarization is computed through
\begin{equation}
{\vec{P}}(E,t)=\frac{\langle\Psi(t)| \vec{s}\delta(E-{\mathcal{H}})+\delta(E-{\mathcal{H}})\vec{s}~|\Psi(t)\rangle}{2\langle\Psi(t)|\delta(E-{\mathcal{H}})|\Psi(t)\rangle}
\label{time-dependence0}
\end{equation}
where $\vec{s}$ are the spin Pauli matrices and $\delta(E-{\mathcal{H}})$ is the spectral measure operator. The time evolution of electronic wavepackets $|\Psi(t)\rangle$ is obtained by solving the Schr\"{o}dinger equation \cite{Roche1997,Roche2014b}, starting from random-phase states $|\Psi(t=0)\rangle=|\varphi_\text{RP}\rangle$ with an initial out-of-plane ($z$ direction) or in-plane polarization ($x,y$ direction).  The random phase states can be generally expressed as $|\varphi_\text{RP}\rangle=
\frac{1}{\sqrt{N}}\sum_{j=1}^{N}  \left(\begin{array}{c}\cos(\theta_{j}/2)\\
 e^{i\Phi_{j}}\sin(\theta_{j}/2)\end{array}\right) e^{2i\pi\alpha_{j}}|j\rangle$, where $(\Phi_{j},\theta_{j})$ gives the spin orientation of orbital $|j\rangle$ in the spherical coordinate, whereas $\alpha_{j}$ is a random number in the $[0, 1]$ interval \cite{Dinh2014}. An energy broadening parameter $\eta$ is introduced for expanding $\delta(E-{\mathcal{H}})$ through a continued fraction expansion of the Green's function \cite{Roche1997,Roche2014b}. An average over few tens of random phases states is usually sufficient to converge the expectation values.
This method has been previously used to investigate spin relaxation in gold-decorated graphene \cite{Dinh2014}, hydrogenated graphene \cite{Soriano2015} and recently, SOC coupled graphene under the effect of electron-hole puddles \cite{DinhScRep}.  

\begin{figure}[hbtp]
\begin{center}	
\includegraphics[width=0.45\textwidth]{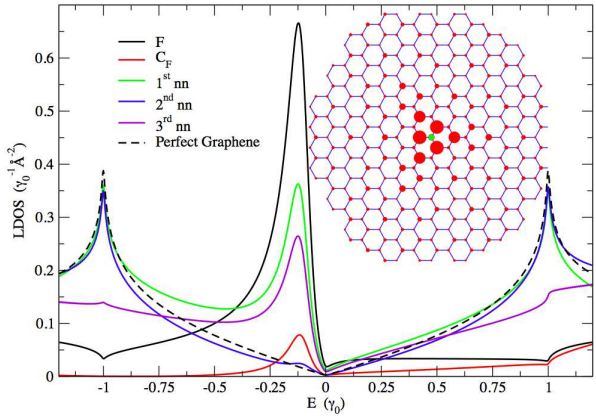}
\caption{Local density of state (LDOS, solid lines) around the fluorinated carbon $C_F$ in the comparison with the pristine graphene one (dashed line). Inset: SPP around the fluorinated carbon (marked by the green circle). The radii of the circle is proportional to the LDOS, the state is projected at the resonant energy $E_R=-0.125\gamma_0$.}
\label{Fig1}
\end{center}
\end{figure}

{\it Impurity resonance and sublattice pseudospin polarization.}  Electronic calculations show that unlike hydrogen, the fluorine adatom is a broad scatterer \cite{Irmer2015}. The density of states (DOS) of a $40\times 40$ supercell (about $0.03\%$) exhibits a resonant peak at 
about $260$ meV \cite{Irmer2015} below the Dirac point. Fig. 1 shows the local density of state (LDOS) on the sites close to the fluorinated carbon $C_F$ using the tight-binding model for ${\cal H}$ (Eq.(1)). All the LDOSs present a strong electron-hole asymmetry with broad peaks at about $E_R=-0.125\gamma_0=-325$ meV which are evidences of fluorine-induced resonant effect. More interestingly, the height of resonant peaks discloses a sublattice asymmetry with less state occupancy on the sublattice related to the fluorinated carbon $C_F$. This is the signature of a SPP which is present in graphene when the A-B symmetry is broken such as in the case of hydrogenated, nitrogen-doped graphene or graphene with vacancies. Fluorine ad-atoms induce a long range pseudospin-polarized region. Fig. 1 (inset) shows the LDOS at the resonant energy $E_R$ (represented by the radius of circles) on more than 300 atoms around the fluorinated carbon (marked by the green dot). At the edge of this area one can still see the difference between  LDOSs on two different sublattices.  Here we will show that the SPP has a direct impact on spin lifetime in fluorinated graphene.

{\it Strong electron-hole asymmetry of the spin lifetime.}  We compute the expectation value of the out-of-plane spin component $P_{z}(E,t)= P_{\perp}(E,t)$  and the in-plane spin component $P_{x}(E,t)= P_{\parallel}(E,t)$ of spin polarization in fluorinated graphene using Eq.(\ref{time-dependence0}). Fig. 2c (inset) shows the evolution of spin polarization $P_z(t)$ at the Dirac point (black and green solid lines) and at the resonant energies $E_R$ (red and blue solid lines) for  $0.01\%$ and $0.02\%$ of fluorine on graphene. There are two interesting features of the spin signals. The first one is the remarkably slow decay of the time-evolution of the spin polarization at the resonance ($E=E_{R}$) compared to that occurring at the Dirac point. The second characteristic is the enhancement of spin polarization when increasing the percentage of fluorine (see illustration in Fig. 2a and Fig. 2b).

Such remarkable features are further manifested in the spin lifetime $\tau_s$ (Fig. 2c, main frame), which are extracted from the spin polarization by fitting the obtained data with an exponential decay $P_{x,z}(t)=P_{x,z}(t_0)e^{-(t-t_0)/\tau_s^{||,\bot}}$ (dashed lines in the inset of Fig. 2c). This fitting is performed starting from the time $t_0=30$ ps to avoid the initially transient fast decay which is usually observed for strong disorder, especially at DP \cite{Soriano2015}. The spin lifetime exhibits a strong electron-hole asymmetry with huge increase of spin lifetime with a maximum close to but not exactly at the resonant energy (about one order of magnitude compared to one in the electron side). 

\begin{figure}[hbtp]
\begin{center}	
\includegraphics[width=0.45\textwidth]{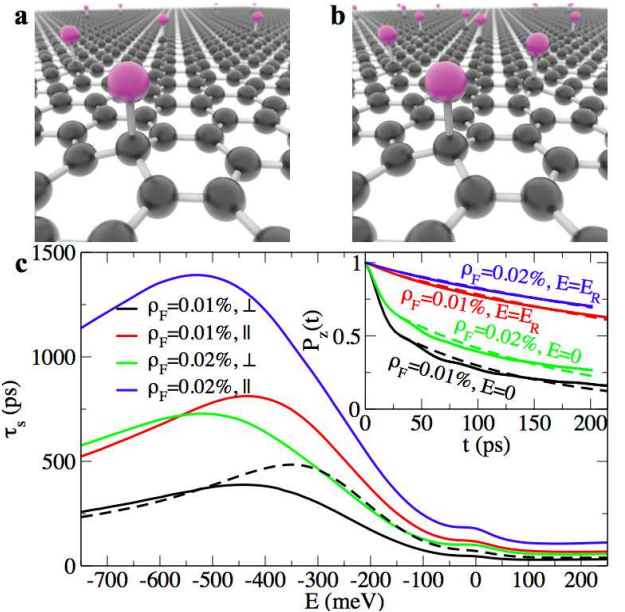}
\caption{(a) and (b) show ball-and-stick models for fluorine functionalized graphene for two different densities. (c) spin relaxation time $ \tau_s$ for in-plane $||$ and out-of-plane $\bot$ spin components for $0.01\%$ and $0.02\%$ of fluorine on graphene (dashed line gives $\tau_s^{\bot}$ for $0.01\%$ fluorine, neglecting the SOC term). Inset: spin polarization evolution at the Dirac point and at resonant energy $E_R$ for $0.01\%$ and $0.02\%$ of fluorine on graphene. }
\label{Fig2}
\end{center}
\end{figure}

The energy of this maximum is shifted to the hole side compared to the resonant energy $E_R$ and this  energy shift increases with the fluorine concentration. This shift is attributed to the SOC effects caused by fluorine ad-atoms. Indeed, turning off the SOC induced by fluorine leads to the spin lifetime (dashed line) with the peak exactly at resonant energy $E_R$.  More remarkably, the increase of fluorine percentage leads to an enhancement of $\tau_s$ (Fig. 2c). This is counterintuitive because fluorine  was predicted to induce a giant SOC in graphene which should lead to a decrease of spin lifetime with fluorine density $\rho_F$ \cite{Irmer2015}.  Actually, the SOC induced by fluorine does not play a major role here. Indeed, in absence of SOC induced by fluorine ($\Lambda_I^B=\Lambda_R=\Lambda_{PIA}=0$), $\tau_s$ shows similar energy dependence (see dashed line in Fig. 2c for $\tau_s^{\perp}$ and $0.01\%$ fluorine atoms neglecting their SOC contribution).  

{\it Dyakonov-Perel mechanism.} Fluorine ad-atoms also induce momentum scattering which yield randomization of the spin precession. This usually leads to a Dyakonov-Perel relaxation mechanism in which the spin lifetime $\tau_s$ is inversely proportional the momentum relaxation time $\tau_p$, i.e. $\tau_s^{||}=2\tau_s^{\perp}=\hbar^2/(2\lambda_R^2\tau_p)$ \cite{DYA_SPSS13,Huertas2009,Ertler2009,Zhang2012}. This scaling can be clearly observed in Fig. 2c where $\tau_s$ upscales with the fluorine density $\rho_F$ almost linearly as expected from a Fermi golden rule argument.  To further confirm the mechanism at play, the momentum relaxation time $\tau_p$ is computed numerically using a real-space order-N approach \cite{Roche2014b}.

Fig. 3a shows the energy dependence of $\tau_p$ for 0.02\% of fluorine on graphene with a minimum close to $E_R$, pinpointing the resonance induced by fluorine (identified by the peak in the LDOS of atoms in the distance of twice carbon bond length from $C_F$, see red dashed line). To further confirm the relaxation mechanism, we compute the product of $\tau_s\tau_p$ (see Fig. 3b). The obtained numerical data (black solid line) close to the resonance are fairly consistent with a Dyakonov-Perel mechanism (red dashed line) up to a factor $\alpha \in [0.6;1.4]$ ($\tau_s^{||}\tau_p=\frac{\alpha\hbar^2}{2\lambda_R^2}$). A final evidence is given by the spin lifetime anisotropy of $\tau_s$ obtained in Fig. 3c. Indeed, the ratio of in-plane and out-of-plane spin lifetimes  $\tau_s^{||}/\tau_s^{\bot}\sim 2$ (within $10\%$ error), well agrees with analytical calculations performed in model systems \cite{Zhang2012}. Some deviation is observed close to the Dirac point, where the spin-pseudospin entanglement effects are maximized.

\begin{figure}[hbtp]
\begin{center}	
\includegraphics[width=0.45\textwidth]{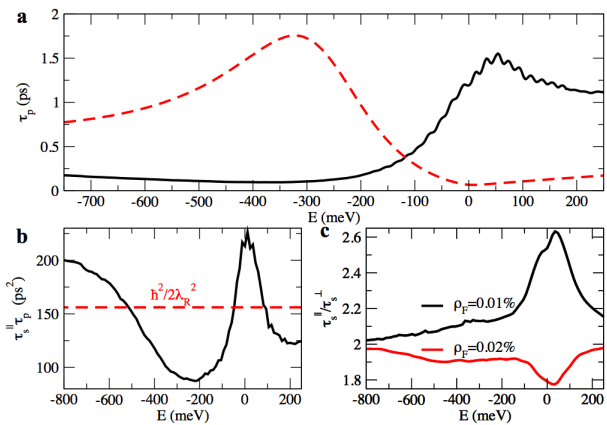}
\caption{(a) Energy dependence of the momentum relaxation time $\tau_p$ for $0.02\%$ of fluorine on graphene. (b) Numerical product of $\tau_s^{||}\tau_p$ compared with the analytical value (dashed line) from ref.\cite{Ertler2009}. (c) The ratio of in-plane and out-of-plane spin relaxation times for $0.01\%$ and  $0.02\%$ of fluorine on graphene.}
\label{Fig3}
\end{center}
\end{figure}

{\it Discussion.} The enhancement and the energy-dependence of $\tau_s$ is a direct consequence of defect-induced sublattice pseudospin polarization (illustrated in Fig. 1, inset).  In supported ultraclean graphene, the Rashba SOC $\lambda_R$ induced by the substrate or the gate voltage dictates the spin dephasing of propagating charges, as shown experimentally \cite{Guimaraes2014}. It is worth mentioning that the spin lifetime caused by this background Rashba SOC is totally electron-hole symmetric \cite{DinhScRep}. On the hole side, the induced SPP around fluorine defects locally suppress the Rashba SOC and consequently enhance spin lifetime up to the range of nanoseconds whereas $\tau_s$ is more strongly reduced on the electron side, with $\tau_s\sim 100$ ps. This phenomenon can be qualitatively understood using both the continuum and tight-binding models.  In the continuum model, the Hamiltonian $\mathcal{ H}_G$ (Eq.(\ref{HamilG})), including spin-orbit interaction, can be approximated as $h_G(\vec{k})=\hbar v_F(\eta \sigma_xk_x+\sigma_yk_y)+\lambda_R(\eta \sigma_xs_y-\sigma_ys_x)+\lambda_I \eta\sigma_z s_z$,  where $\sigma$ and $s$ are Pauli matrices representing the sublattice pseudospin and spin degrees of freedom respectively, while $\eta=1(-1)$ corresponds to the K (K') valley (here intervalley coupling is neglected in the discussion). The magnitude of the Rashba magnetic field is proportional to the in-plane component of pseudospin ($\sigma_x, \sigma_y$) \cite{Dinh2014}
which is reduced by approaching the area around fluorine due to the formation of SPP. The reduction of local effective Rashba magnetic field entails the enhancement of spin lifetimes which is maximum close to the resonant energy $E_R$ where the SPP is maximum. In the tight-binding model, the peculiar sublattice occupancy of impurity states gives rise to an increase of the next-nearest-neighbor hopping probability (intrinsic SOC) and a decrease of the nearest-neighbor hopping probability (Rashba SOC), which being the main factor for spin relaxation, also explains the spin lifetime enhancement. 

One notes that SPP is not unique to fluorine adsorption in the weak density limit, but can be also generated by nitrogen substitutions \cite{Zhao2011}, grafted molecules \cite{Mali2015}, hydrogen ad-atoms or any other effect breaking A-B sublattice symmetry. However the unveiled phenomenon of electron-hole spin transport asymmetry should be maximized in absence of magnetic moments, which disfavor long spin propagation \cite{Kochan2014,Soriano2015}. Besides, in contrast to the Elliot-Yafet mechanism predicted for magnetic impurities \cite{Kochan2014,Soriano2015,Bundesmann2015}, ad-atoms such as fluorine are here shown to entail a Dyakonov-Perel mechanism, in agreement with many experiments on functionalized graphene \cite{Wojtaszek2013,Swartz}. We note that the considered dilute fluorine limit is accessible experimentally \cite{Hong2011,Avsar2015}, and that chemical bonding of fluorine ad-atoms is theoretically tunable with electric field \cite{Sofo2011,Guzman-Arellano2014,Usaj1}, a fact which could help controlling the level of adsorption and the possibility to switch on and off the spin transport asymmetry generated by impurities. Finally we observe that spin dynamics could be a smoking gun for unveiling new quantum phase transition resulting from the competition between different ground states (such as those characterized by spin-degenerate and magnetic bound states \cite{Guessi2015}), or to scrutinize the origin of the saturation of coherence times in weak localization measurements \cite{Hong2012}.

{\it Acknowledgements}. This work has received funding from the European Union Seventh Framework Programme under grant agreement 604391 Graphene Flagship. S.R. acknowledges the Spanish Ministry of Economy and Competitiveness for funding (MAT2012-33911), the  Secretaria de Universidades e Investigacion del Departamento de Economia y Conocimiento de la Generalidad de Catalu\~{n}a and the Severo Ochoa Program (MINECO SEV-2013-0295).

\bibliography{biblio}

\end{document}